\documentclass[ aip,jap, jmp,amsmath,amssymb, reprint,]{revtex4-1}
\usepackage{graphicx}
\usepackage{dcolumn}
\usepackage{bm}
\usepackage{natbib}

\begin{document}
\preprint{AIP/123-QED}
\title{Rapid diffusion of electrons in GaMnAs}
\date{\today}
\author{C. P. Weber}\email{cweber@scu.edu}
\author{Eric A. Kittlaus}
\author{Kassandra B. Mattia}
\author{Christopher J. Waight}
\affiliation{Department of Physics, Santa Clara University, 500 El Camino Real, Santa Clara, CA 95053-0315, USA}
\author{J. Hagmann}
\author{X. Liu}
\author{M. Dobrowolska}
\author{J. K. Furdyna}
\affiliation{Department of Physics, University of Notre Dame, Notre Dame, IN 46556, USA}

\begin{abstract}
We report ultrafast transient-grating measurements, above and below the Curie temperature, of the dilute ferromagnetic semiconductor (Ga,Mn)As containing 6\% Mn. At 80 K (15 K), we observe that photoexcited electrons in the conduction band have a lifetime of 8 ps (5 ps) and diffuse at about 70 cm$^2$/s (60 cm$^2$/s). Such rapid diffusion requires either an electronic mobility exceeding 7,700 cm$^2$/Vs or a conduction-band effective mass less than half the GaAs value. Our data suggest that neither the scattering rate nor the effective mass of the (Ga,Mn)As conduction band differs significantly from that of GaAs.
\\ \\
The following article appeared in \textit{Applied Physics Letters} and may be found at\\ \url{http://link.aip.org/link/?APL/102/182402}
\\
(This version of the article differs slightly from the published one.) 
 \\ \\
\textit{Copyright 2013 American Institute of Physics. This article may be downloaded for personal use only. 
Any other use requires prior permission of the author and the American Institute of Physics.}
\end{abstract}
\maketitle

\section{Introduction}

The dilute magnetic semiconductor (Ga,Mn)As is considered an archetypal spintronic material. Despite years of intensive study, the band structure of its holes continues to excite debate.\protect{\cite{SamarthNMat2012}} The conduction band, on the other hand, is nearly unstudied. Its transport properties, however, bear importantly on the proposed ``magnetic bipolar transistor,''\protect{\cite{Fabian2002, Flatte2003, Lebedeva2003}} which would operate through the diffusion current of electrons in a (Ga,Mn)As base. The device's current gain and spin injection would both increase with increasing minority-carrier diffusivity.

We are not aware of measurements of electron transport in (Ga,Mn)As, and the extensive study of the valence band offers only mixed guidance as to what might be expected of the conduction band. Recent experiments indicated a long scattering time\protect{\cite{Ohya2011}} and GaAs-like effective mass\protect{\cite{Ohya2011, ChaplerPRB2012}} in the valence band. On the other hand, the hole mobility of (Ga,Mn)As is typically low,\protect{\cite{Jungwirth2007}} likely due to disorder resulting from Mn doping, and the mean free path is estimated as just a few nanometers.\protect{\cite{MocaPRL2009}} The Mn-induced disorder has been imaged by scanning tunneling microscopy, and persists from the valence band through the conduction-band edge,\protect{\cite{Richardella2010}} which suggests a short scattering time for electrons.

Here we measure that diffusion of photoexcited electrons, during the first several picoseconds after excitation, is about 70 cm$^2$/s. Such high diffusivity implies that the electrons' mobility must be high, or their effective mass low, or both. In particular, our data are consistent with an electron effective mass equal to that of GaAs and a mobility greater than 7,700 cm$^2$/Vs.

\section{Methods}

Our (Ga,Mn)As sample was grown by molecular beam epitaxy on a substrate of (001) semi-insulating GaAs. The structure consists of a 250 nm stop-etch layer of Al$_{0.35}$Ga$_{0.65}$As, a 25 nm buffer of GaAs, and the 800 nm sample of Ga$_{0.94}$Mn$_{0.06}$As, grown with a Mn cell temperature of 760 $^{\circ}$C and a substrate temperature of 215 $^{\circ}$C. The film's resistivity indicates a Curie temperature $T_C\approx56$ K. From these data we estimate the hole density as $p_0\approx1\times10^{20}$ cm$^{-3}$. After growth the film was affixed to a sapphire window and the GaAs substrate removed by chemical etching.\protect{\cite{KimJVSTB1998}} Since Al$_{0.35}$Ga$_{0.65}$As is transparent at our measurement wavelength of 810 nm, our signal arises entirely in the (Ga,Mn)As film. The thickness of the (Ga,Mn)As film was chosen to ensure adequate signal.\protect{\cite{Yildirim2011}} We expect a high concentration of interstitial Mn defects, as films this thick cannot be effectively annealed.

To measure electron diffusion we use an ultrafast transient-grating method.\protect{\cite{Eichler1986}} A pair of ``pump'' laser pulses, split by a diffractive optic,\protect{\cite{Goodno1998,Maznev1998}} are simultaneously incident on the sample. The two pulses are non-collinear and interfere; their absorption excites photocarriers in a sinusoidal pattern with wavelength $\Lambda$ and wavevector $q=2\pi/\Lambda$. By locally modifying the index of refraction, the carriers create a ``grating'' off of which time-delayed probe pulses diffract. As carriers recombine and diffuse, the grating amplitude decays at a rate
\begin{equation}\frac{1}{\tau(q)}=D_aq^2+\frac{1}{\tau_0}.
\label{DiffusionEq}
\end{equation}
$D_a$ is the ambipolar diffusion coefficient and $\tau_0$ is the lifetime of spatially uniform excitation. Measurement at several $q$ determines $D_a$. Coulomb attraction constrains the electrons and excess holes to diffuse together, making diffusion ambipolar:\protect{\cite{vanRoosbroeck1953}} the motion is controlled primarily by the species with the lower conductivity.

We measure the diffracted probe amplitude in a reflection geometry, improve measurement efficiency by heterodyne detection,\protect{\cite{Vohringer1995}} and suppress noise by 95-Hz modulation of the grating phase and lock-in detection.\protect{\cite{Weber2005}} The laser pulses have wavelength near 810 nm, duration 120 fs, and repetition rate 80 MHz. The pump pulses have fluence at the sample of 4.2 $\mu$J/cm$^2$, and the probe pulses 0.35 $\mu$J/cm$^2$. At 810 nm GaAs has an absorption length\protect{\cite{SturgePhysRev1962}} of order 1 $\mu$m and reflectivity of 0.3, so each pair of pump pulses photoexcites electrons and holes at a mean density of $n_{\text{ex}}\approx 1.2\times10^{17}$ cm$^{-3}$. The incident photons have energy about 30 meV greater than the bandgap of GaAs at 80K. (Ga,Mn)As is known to suffer bandgap narrowing of order 120 meV,\protect{\cite{OhnoPhysicaE2002}} so we estimate that our photoexcited electrons have excess energy of 150 meV. Screening of the electron-LO phonon interaction\protect{\cite{DasSarmaPRB1987}} could slow the electrons' cooling to the lattice temperature, but they will rapidly thermalize with the much more numerous holes through carrier-carrier scattering. The holes' specific heat is about 100 times greater than the electrons', resulting in a total heating of just 20 K. Thus the rapid diffusion that we observe does not reflect the transport properties of hot electrons, but of equilibrium ones.\protect{\cite{temperaturenote, nonsinusoidalnote}}

\begin{figure}
\includegraphics[width=3.45 in]{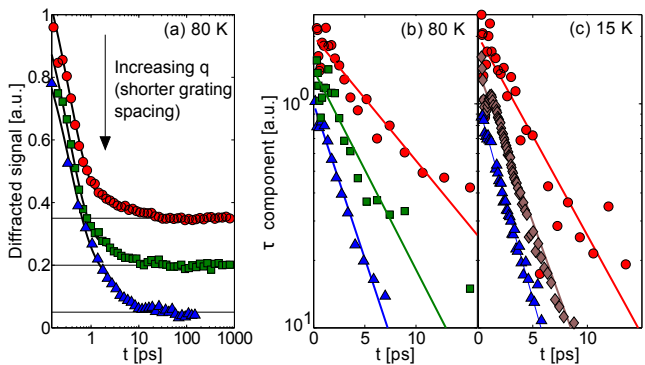}
\caption{(Color online) Diffracted amplitude $S(t)$ for q=12,600 cm$^{-1}$ (circles), 37,700 cm$^{-1}$ (squares), 45,300 cm$^{-1}$ (diamonds), and 52,800 cm$^{-1}$ (triangles). Scaled and offset vertically for clarity. Solid lines are fits to the form of Eq. \ref{doubleexp}. (a) Entire signal; semilog $t$. (b,c) Signal with constant and $\tau_\text{fast}$ terms subtracted, leaving just the $\tau(q)$ component; semilog $S$.}
\label{Curves}
\end{figure}

\begin{figure}
\includegraphics[width=3.45 in]{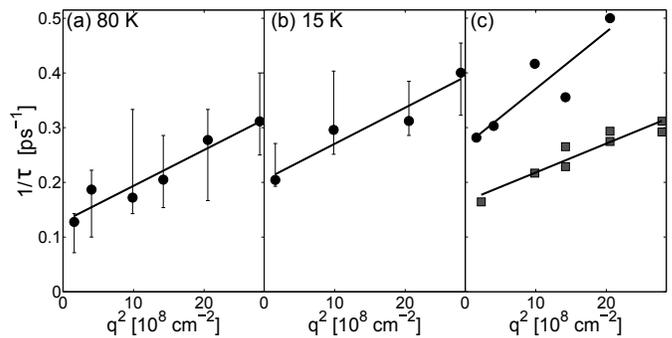}
\caption{(a,b) Decay rate of the signal's $\tau(q)$ component \textit{vs}. $q^2$. Points and error bars are obtained by procedures described in the text. The lines are least-squares fits of the points to Eq. \ref{DiffusionEq}, revealing diffusion of photoexcited carriers. (c) Similar plots for a 5\%-Mn, non-etched sample at 15 K (squares) and 80 K (circles). Errors are smaller than for the etched sample and can be estimated by the scatter of the points.}
\label{Diffusionplot}
\end{figure}

\section{Results} 

Figure \ref{Curves}a shows the diffracted signal $S(t)$ as a function of time for gratings of several $q$. In each case the signal is well described by the form
\begin{equation}S(t)=Ae^{-t/\tau_\text{fast}}+Be^{-t/\tau(q)}+C,
\label{doubleexp}\end{equation}
with $\tau_\text{fast}\approx0.34$ ps, $2.5\text{ ps}<\tau(q)<8$ ps, $B/A=0.11\pm.06$, and $C/A=-0.012\pm.007$ (The errors are standard deviations.) Of these three components of our signal, we will see below that the $\tau(q)$ term reveals the dynamics of free electrons diffusing  at about 70 cm$^2$/s. 

Previous ultrafast experiments on (Ga,Mn)As have observed signals with a very slow component corresponding to our $C$ term, and attributed it to trapped carriers\protect{\cite{ZahnJAP2010, KimPSSC2005, KimAPL2006}} or to residual heat from the excitation pulse.\protect{\cite{KojimaPRB2003, KojimaPRB2007}} Such experiments have also observed fast processes on the scale of $\tau_\text{fast}$,\protect{\cite{ZahnJAP2010, KimPSSC2005, KimAPL2006} which were variously attributed to intraband relaxation\protect{\cite{KimPSSC2005}} or trapping\protect{\cite{ZahnJAP2010}} of electrons at As antisite defects. Rapid trapping is consistent with our observation of electrons' diffusion at later times: the number of electrons photoexcited likely exceeds the number of trapping sites, leaving some electrons untrapped and free to diffuse. Our observations contradict reports\protect{\cite{KojimaPRB2003}} that \textit{all} carriers are trapped and recombined within a picosecond. 

Since $D_a$ is determined from $\tau(q)$, it is critical to find the latter values reliably, despite the constant and $\tau_\text{fast}$ terms, which introduce free parameters when fitting the data to Eq. \ref{doubleexp}. By measuring $S(t)$ at times much greater than $\tau(q)$, we determine $C$ unambiguously. Since $\tau_\text{fast}$ appears to have similar values at all $q$, we reduce parameters further by fitting all the data with the single value $\tau_\text{fast}=0.34$ ps. The resulting values of $1/\tau(q)$ appear as the points in Fig. \ref{Diffusionplot}a,b. This procedure gives reliable fits to the data's $\tau(q)$ component, as evidenced by Fig. \ref{Curves}b,c, in which we plot the measured $S(t)$ with the constant and $\tau_\text{fast}$ terms subtracted. The signal decays exponentially, and the decay is visibly faster at high $q$, indicative of diffusive motion. Finally, we repeat each fit for a series of \textit{fixed} values of $\tau(q)$ and judge ``by eye'' for what range of $\tau(q)$ the fits appear to plausibly reproduce the data. The outer limits of this range determine the error bars shown in Fig. \ref{Diffusionplot}a. 

The line in Fig. \ref{Diffusionplot}a is a fit to the form of Eq. \ref{DiffusionEq}, giving $D_a=67$ cm$^2$/s and $\tau_0=8$ ps. (Fits to the lower or upper ends of the error bars give values of $D_a$ ranging from 60 to 80 cm$^2$/s, and $\tau_0$ from 14 to 6 ps.) For the 15-K data (Fig. \ref{Diffusionplot}b), $D_a$ is between 50 and 70 cm$^2$/Vs and $\tau_0$ is between 3.4 and 5.2 ps. This figure makes it evident that our signal's $\tau(q)$-component arises from ambipolar diffusion of free electrons, for the diffusivity is much faster than expected for heat or trapped carriers, while being comparable to prior measurements in $p$-type GaAs\protect{\cite{Akiyama1994, Linnros1994, Paget2012}} and InP.\protect{\cite{Weber2013}}

We note that we see much the same result on a sample with nominal Mn doping of 5\% and 800-nm thickness, which gives $D_a$ of 53 cm$^2$/s at 15 K and 104 cm$^2$/s at 80 K (Fig. \ref{Diffusionplot}c). The temperature dependence may arise from the electrons' mobility, or from spin-splitting of the conduction band below $T_C$. Our etching procedure succeeds rarely; most often the sample disappears completely or the etched sample scatters too much light to allow measurements. For this reason the 5\% sample is not etched, and the measured signal may include a small component from the GaAs substrate. Thus these measurements should be seen merely as corroboration of our result from the etched, 6\% sample, which shows similar $D_a$ and similar variation with temperature.

Though the $\tau_0=8$ ps that we measure at 80 K may seem too short to be the free-electron lifetime, it is consistent with previous observations in low-temperature-grown GaAs\protect{\cite{Loka1998, Stellmacher2000}} and in (Ga,Mn)As,\protect{\cite{MitsumoriPRB2004, KimPSSC2005, SunJAP2006, RozkotovaAPL2008, LiuAPS2008}} where a transient lasting 6 to 14 ps was attributed to non-radiative recombination of trapped carriers.\protect{\cite{SunJAP2006, RozkotovaAPL2008}} Our signal arises from free electrons, not trapped, but is sensitive to recombination because, when all traps are filled, free electrons are trapped only as quickly as recombination empties traps. In what follows, $n_\text{f}$, the number of \textit{untrapped} electrons, will be important. Our ``best guess'' is that $n_\text{f}/n_\text{ex}\approx B/A\approx0.1$, \textit{i.e.}, that the relative amplitude of the $\tau(q)$ and $\tau_\text{fast}$ terms represents the proportion of photoexcited electrons remaining untrapped after 0.34 ps. Note, however, that our conclusions will hold true for the entire possible range ${0<n_\text{f}<n_\text{ex}}$.

\begin{figure}
\includegraphics[width=3.45 in]{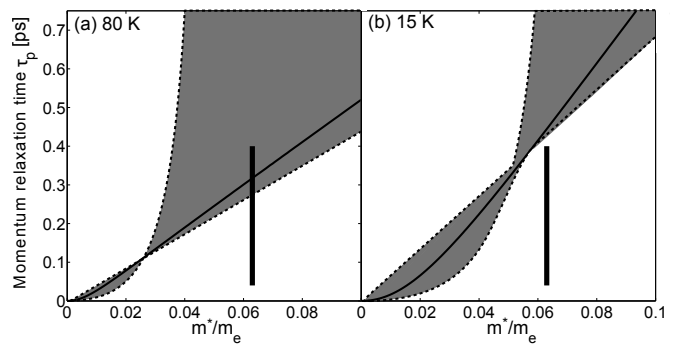}
\caption{For the simple model described, the relation between the electrons' effective mass $m^*/m_e$ and their momentum-relaxation time $t_p$, as constrained by our data. The solid curve assumes a density of untrapped carriers $n_\text{f}=n_\text{ex}/10$. The shaded region is bounded by $n_\text{f}=n_\text{ex}$ (curved) and $n_\text{f}=0$ (diagonal). The mobility exceeds 7,700 cm$^2$/Vs for all points above the diagonal boundary. The thick line is the range of values expected for GaAs at 80 K.}
\label{mvstau}
\end{figure}

\section{Discussion}

We now show that it is possible to use our measurements of $D_a$ and resistivity $\rho$ (of the holes)\protect{\cite{resistivitynote}} to infer conduction-band properties, despite our uncertain knowledge of three key parameters. The first is $\mu_e$, the electrons' mobility; second is $n_\text{f}$. Finally, the electrons' charge-susceptibility is ${\chi_e=dn/d\mu_{\text{chem}}}$, with $\mu_{\text{chem}}$ the chemical potential. This quantity approaches $n_\text{f}/k_BT$ at high temperature, and at low temperature approaches $N_e$, the electrons' density of states at their quasi-Fermi energy; it thus provides insight into the electrons' effective mass $m^*$. (With $\chi_h$ defined similarly for holes, the reduced susceptibility is ${1/\chi^*\equiv1/\chi_e+1/\chi_h}$. In this experiment $\chi_h>\chi_e$ and $\chi^*\approx\chi_e$; we nonetheless include the hole susceptibility in our numerical calculations.\protect{\cite{holemassnote}})

By the Einstein relation, the electrons' diffusivity is
\begin{equation}D_e=\frac{n_\text{f}\mu_e}{e\chi_e},
\end{equation}
and similarly for holes. The photoexcited electrons and holes must diffuse together, giving ambipolar diffusion\protect{\cite{vanRoosbroeck1953}} ${D_a=(\sigma_eD_h+\sigma_hD_e)/(\sigma_e+\sigma_h)}$, from which
\begin{equation}\frac{1}{D_a}=\left(e^2\rho+\frac{e}{n_{\text{f}}\mu_e}\right)\chi^*.
\label{inverseDa}
\end{equation}
Here we have used ${n_{\text{ex}}\ll p_0}$. Our measurements suffice to specify neither $\chi^*$ nor $\mu_e$, but do constrain the relation between them: Eq. \ref{inverseDa} shows that the larger the mobility, the larger $\chi^*$. 

To further explore the relationship between $\mu_e$ and $\chi^*$, it will be instructive to recast it in terms of $m^*$ and the momentum-relaxation time $t_p$, and to compare our results on (Ga,Mn)As with the known properties of GaAs (which has $m^*/m_e=0.063$). In undoped GaAs the mobility at 80 K can reach $2\times10^5$ cm$^2$/Vs, but for $n\approx n_{\text{ex}}$ the mobility should be about $10^3$ to $10^4$ cm$^2$/Vs,\protect{\cite{Rode1975_2, Lowney1991}} and comparable at 15 K.\protect{\cite{Stillman1970}} The latter values correspond to momentum relaxation times $t_p$ of 0.04 to 0.4 ps. The range of typical GaAs parameters appears as a thick, vertical lines in Fig. \ref{mvstau}a,b, where we choose to plot $t_p$ rather than $\mu_e$ since mobility contains both $t_p$ and $m^*$. 

Turning back to our (Ga,Mn)As results, at finite temperature $\chi^*(m^*)$ must be calculated numerically.\protect{\cite{Mohankumar1995} For this purpose we introduce a simple model that treats both bands as isotropic and parabolic, the hole mass\protect{\cite{holemassnote}} as $0.53m_e$, and the the conduction-band effective mass $m^*$ as a free parameter. We calculate first $\chi^*(m^*)$, then the corresponding $\mu_e(\chi^*,D_a,\rho)$, and finally the momentum-relaxation time ${t_p=\mu_em^*/e}$; the results for 80 K appear in Fig. \ref{mvstau}a. Since $n_\text{f}$ is unknown, we repeat the calculation for a range of $n_\text{f}$: the solid curve assumes $n_\text{f}/n_\text{ex}=0.1$, and we shade the entire possible region ${0<n_\text{f}/n_\text{ex}<1}$. Note that all points on the diagonal line for $n_\text{f}=0$ have $\mu_e=7,700$ cm$^2$/Vs, and that points in the allowed region above the line have yet higher mobility. We believe, therefore, that $\mu_e\geq7,700$ cm$^2$/Vs; this value is high, but not without precedent in GaAs.

Figure \ref{mvstau}b shows the results of a similar calculation for our 15-K data. For the region above the diagonal, $\mu_e\geq 12,000$ cm$^2$/Vs. The calculation assumes spin-splitting of the conduction band; a calculation without spin splitting would give the same lower bound on $\mu_e$ but would shift the curved boundary (corresponding to  ${n_\text{f}=n_\text{ex}}$) leftward. Within the experimental uncertainties, the 15-K mobility is similar to that at 80 K. 

It may seem that (Ga,Mn)As, as a disordered material, ought to have mobility well below that of GaAs. Indeed, our experimental results do not, strictly, exclude a low mobility: it is allowed if $m^*/m_e<0.03$ (Fig. \ref{mvstau}a, lower left). We are not, however, aware of any mechanism to suppress the conduction-band density of states so strongly. It would also seem unlikely for $t_p$ to greatly exceed 0.4 ps, which suggests that $m^*$ cannot much exceed the GaAs value. Thus the most plausible scenario supported by our data is that neither the conduction band's scattering rate nor its effective mass is much influenced by Mn doping. We note that some recent results indicated that the (Ga,Mn)As \textit{valence} band has GaAs-like effective mass\protect{\cite{Ohya2011, ChaplerPRB2012}} and a low scattering rate\protect{\cite{Ohya2011}}. Indeed, those results are even more surprising than ours, because Mn states are much closer to the valence band edge than to the conduction band.

The high diffusivity we measure could aid the operation of a GaMnAs-based magnetic bipolar transistor.\protect{\cite{Fabian2002, Flatte2003, Lebedeva2003}} In order to exhibit high current gain, the device needs a minority-carrier diffusion length that greatly exceeds the base width $w$. The electrons' low-temperature diffusion length of $L=\sqrt{D_e\tau_0}\approx170$ nm is not long, but is long enough to allow $w<L/8$ to be achieved by commercially-available photolithographic techniques. If the electrons' spin diffusion $D_s$ is close to $D_e$, such a device would inject a spin polarization proportional to $D_e$ into the collector.\protect{\cite{Fabian2004}} Just as our measurement photoexcites enough electrons to fill the traps, the transistor would need to operate in a high-injection regime. High injection mitigates carrier trapping, and may also be responsible for the electrons' high mobility: an As$^+$ antisite becomes neutral when trapping an electron.

\section{Conclusions}

In conclusion, we performed transient-grating experiments to measure the diffusion of photoexcited electrons in Ga$_{0.94}$Mn$_{0.06}$As. We found a signal consisting of three components; the component revealing motion of free electrons had a diffusivity of 60-80 cm$^2$/s at 80 K and 50-70 cm$^2$/s at 15 K. This rather high diffusivity, along with the measured resistivity, constrains the electrons to have either a high mobility or very a low effective mass. We consider it most plausible that the electrons' effective mass and mobility are both comparable to those of GaAs, with $\mu_e\geq7,700$ cm$^2$/Vs. The mobility is nonetheless surprisingly high, given the disorder in (Ga,Mn)As alloys, and reflects favorably on the possibility of GaMnAs-based magnetic bipolar transistors. Our result suggests the value of further measurements, such as a Shockley-Haynes experiment, that would measure ambipolar \textit{mobility} and would thus allow the determination of both the electrons' mobility and their effective mass. 

\begin{acknowledgements}
This work was supported by the National Science Foundation Grants No. DMR-1105553 and DMR 10-05851. E.A.K. and K.B.M. were partly supported by Santa Clara University's Hayes and Clare Booth Luce Scholarships, respectively.
\end{acknowledgements}


\begin{thebibliography}{10}
\bibitem{temperaturenote}
In calculations we incorporate this heating by using electron temperature of 35
  K (100 K) for measurements at lattice temperature of 15 K (80 K).
\bibitem{nonsinusoidalnote}
Diffusivity varies with $n_{\text{ex}}$, and the peaks of our grating will
  diffuse more rapidly than the troughs. Our numerical simulations indicate
  that for the densities and temperature used in this experiment, this effect
  can cause us to underestimate $D_a$ by up to 30\%, making electron mobility
  potentially somewhat higher, but not lower, than the values we report.
\bibitem{holemassnote}
$\chi_h$ depends on $p_0$ and the hole density of states, neither of which is
  accurately known. However, the effect of the holes on $\chi^*$ is less than
  8\% provided $p_0$ exceeds 10$^{18}$ cm$^{-3}$ and the hole effective mass
  exceeds 1/20 of the free-electron mass.
\end{thebibliography}
\end{document}